\documentclass[12pt]{article} \textheight=24 true cm
\usepackage{subfigure}
\usepackage{graphicx}
\usepackage{amsmath}
\usepackage{amssymb}
\usepackage{color}
\usepackage[colorlinks]{hyperref}

\begin{document}
\begin{center}
{\Large\bf FRW type of cosmology with a Chaplygin gas }\\[15 mm]
D. Panigrahi\footnote{ Sree Chaitanya College, Habra 743268, India
\emph{and also} Relativity and Cosmology Research Centre, Jadavpur
University, Kolkata - 700032, India , e-mail:
dibyendupanigrahi@yahoo.co.in }
  and S. Chatterjee\footnote{ IGNOU Convergence
Centre, New Alipore College, Kolkata - 700053, India \emph{and
also} Relativity and Cosmology Research Centre, Jadavpur
University,
Kolkata - 700032, India, e-mail : chat\_sujit1@yahoo.com\\Correspondence to
: S. Chatterjee} \\[10mm]

\end{center}

\begin{abstract}
The evolution of a universe modelled as a mixture of generalised
Chaplygin gas and ordinary matter field is studied for a Robertson
Walker type of spacetime. This model could interpolate periods of
a radiation dominated, matter dominated and a cosmological
constant dominated universe. Depending on the arbitrary constants
appearing in our theory the instant of flip changes. Interestingly
we also get a bouncing model when the signature of one of the
constants changes. The velocity of sound may become imaginary
under certain situations pointing to a perturbative state and
consequently the  possibility of  structure formation. We also
discuss the whole situation in the backdrop of wellknown
Raychaudhury equation and a comparison is made with the previous
results.
\end{abstract}

\bigskip
\section{ Introduction}
Three discoveries in the last century have radically changed our
understanding of the universe - as opposed to the idea of
Einstein's static universe Hubble and Slipher(1927) showed that it
is expanding. Secondly CMBR as also primordial nucleosynthesis
analysis in the sixties point to an initial hot dense state of the
universe, which has been expanding for the last 13.5 Gyr. Finally,
if we put faith in Einstein's theory and FRW type of model then as
standardised candles type Ia supernova suggest~\cite{res} that the
universe is undergoing accelerated expansion with baryonic matter
contributing only five percent of the total energy budget. Later
data from CMBR probes~\cite{spe} also point to the same finding.
This has naturally led a vast chunk of cosmology community to
embark on a quest to attempt to explain the cause of the apparent
acceleration. The vexed question in this field is the possible
identification of the processes likely to be responsible for
triggering the late inflation. Researchers are plainly divided
into two broad groups - either modification of the original
Einstein's theory or introduction of any exotic type of fluid like
a cosmological constant or a quintessential type of scalar field.
But the popular explanation with the help of a cosmological
constant is beset with serious theoretical problems because
absence of acceleration at redshifts $z\geq1$ implies that the
required value of the cosmological constant is approximately 120
orders of magnitude smaller than its natural value in terms of
Planck scale~\cite{cop}. As for the alternative quintessential
field~\cite{sam}  we do not in fact have a theory that would
explain, not to mention predict, the existence of a scalar field
fitting the bill without violating the realistic energy
conditions. Moreover we can not generate this type of a scalar
field from any basic principles of physics. Other alternatives
include k-essence~\cite{sch}, tachyon~\cite{gib},
phantom~\cite{eli} and quintom~\cite{guo}. So there has been a
resurgence of interests among relativists, field theorists,
astrophysicists and people doing astroparticle physics both at
theoretical and experimental levels to address the problems
emanating from the recent extra galactic observations without
involving any mysterious form of scalar field by hand but looking
for alternative approaches  based on sound physical principles.
Alternatives include, among others, higher curvature theory,
axionic field and also Brans- Dicke field. Some people attempted
to look into the problem from a purely geometric point of view -
an approach more in line with Einstein's spirits. For example,
Wanas~\cite{wan} introduced torsion while
Neupane~\cite{neu1,neu11} modifies the spacetime with a warped
factor in 5D spacetime in a brane like cosmology and finally
addition of extra spatial dimensions in physics as an offshoot of
prediction from the string theory~\cite{dp1,dp11, sah,kac,car}.
While this torsion inspired inflation has certain desirable
features the problem with Wanas' model is that the geometry is no
longer Riemannian. Further a good number of
people~\cite{kra,aln,sc1,hir} have done away with the concept of
homogeneity itself and have argued that accelerating model and
consequent introduction of exotic matter field have to be invoked
only in FRW type of cosmology. In Tolman Bondi like inhomogeneous
model the apparent dimming of the signals may be explained as a
consequence
of inhomogeneous distribution of matter.\\
\\
While the above mentioned alternatives to explain away the
observed acceleration of the current phase have both positive and
negative aspects the one that caught the attention of a large
number of workers is the introduction of a Chaplygin type of gas
as new matter field to simulate a sort of dark energy. The form of
the matter field is later generalised through the addition of an
arbitrary constant as exponent over the mass density and is
generally referred to as generalised chaplygin
gas(GCG)~\cite{bent,gor1}. Though it suffers from the serious
disqualification that it violates the time honoured principle of
energy conditions its theoretical conclusions are found to be in
broad agreement with the observational results coming out of
gravitational lensing or recent CMBR and SNe data in varied cosmic
probes ~\cite{fabris,col}. This is generally achieved through a
careful maneuvering of the value of the newly introduced arbitrary
constant. To further fine tune the match between the theory and
the very recent observational fallouts the GCG is again modified
via the addition of an ordinary matter field, which is termed in
the literature as modified  chaplygin gas(MCG)~\cite{bena,deb1}.
The viability of such scenarios has been tested by a number of
cosmological probes, including SNe Ia data~\cite{fabris,col},
lensing statistics~\cite{dev1,sil,dev2}, age-reshift
tests\cite{alcaniz}, CMB measurements~\cite{bento}, measurements
of X Ray luminosity of galaxy clusters~\cite{cunha}, statefinder
parameters~\cite{sahni}. In our previous work~\cite{dp2} we have
studied Chaplygin gas model in inhomogeneous space time.  In the
present work we have revisited the dynamics of the FRW model
taking MCG as matter field and have tried to discuss some as yet
unexplored region and have got some interesting results. We have
organised the paper as follows: In section 2 the mathematical
formulation is given and we have ended up with a hypergeometric
solution and also an effective equation of state as $\rho =
\mathcal{W}(t)p$ in section 3. So depending on initial conditions
our model mimics both $\Lambda CDM$ and quiessence models and the
evolution is also shown graphically. We have also made some
detailed discussion on acoustic wave in our model and find that
all possibilities like less/greater than light velocity and even
imaginary values exist in our model. Relevant to mention that
imaginary sound velocity is not that much discouraging in this
context because it gives rise to perturbation and consequent
structure formation~\cite{fab1}. The interesting thing in our
analysis is that we have taken the first order approximation of
the field equation as key equation and subsequently found out the
exact solutions. We are not aware of attempts of similar kind in
the past literature. Moreover it is also found that if an
arbitrary constant appearing in our solution be taken negative the
cosmology bounces back from a minimum. We have also made a
detailed analysis of flip time both analytically and graphically
in this section. In section 4 these conclusions are checked in the
framework of well known Raychaudhury equation. The paper ends with
a discussion in section 5.

\section{ Field Equations}
We consider a spherically symmetric homogeneous spacetime given by
\begin{equation}\label{a}
  ds^{2}= dt^{2}- a^2(t)~(dr^{2}+r^{2}d\Omega^{2})
\end{equation}

where the scale factor, $a(t)$  depends on time only.

 \vspace{0.1 cm}
 A comoving coordinate system is taken
such that $ u^{0}=1, u^{i}= 0 ~(i = 1, 2,3)$ and $g^{\mu
\nu}u_{\mu}u_{\nu}= 1$ where $u_{i}$ is the 4- velocity. The
energy momentum tensor for a dust distribution in the above
defined coordinates is given by

\begin{equation}
T^{\mu}_{\nu} = (\rho + p)\delta_{0}^{\mu}\delta_{\nu}^{0} -
p\delta_{\nu}^{\mu}
\end{equation}

where $\rho(t)$ is the matter density  and $p(t)$ the isotropic
 pressure.

The independent field equations for the metric (1) and the energy
momentum tensor (2) are given by

\begin{equation}
3\frac{\dot{a}^2}{a^2} = \rho
\end{equation}
\begin{equation}
 2 \frac{\ddot{a}}{a} + \frac{\dot{a}^2}{a^2} = - p
\end{equation}

 \vspace{0.2 cm}

From the the Bianchi identity we get for the homogeneous model the
conservation law

\begin{equation}
\nabla_{\nu}T^{\mu \nu}= 0
\end{equation}

which, in turn, yields

\begin{equation}
\dot{\rho} + 3 \frac{\dot{a}}{a} (\rho + p) = 0
\end{equation}

At this stage we assume that we are dealing with a Modified
Chaplygin type of gas (MCG) obeying an equation of state

\begin{equation}
p = A\rho -\frac{B}{\rho^\alpha}
\end{equation}
where $A, B$~ and ~$\alpha$ are constants. The exponent $\alpha$,
from most observational constraints, hovers around unity
~\cite{lima} and the constant $ A$ ranges from $1/3 $ to zero.
Similarly the positive definite constant $B$ is also not exactly
arbitrary. In the equation (7) when the last two terms start to be
of the same order of magnitude the pressure vanishes. In this case
the fluid has pressureless density$\rho_{0}$ corresponding to some
cosmological scale  $a_{0}$ given by $ \rho_{0} = \rho^{\alpha +
1}(a_{0}) = \frac{B}{A}$. Many variants of Chaplygin gas model
have come up in the literature and the equation (7) refers to what
is generally known as the Modified Chaplygin gas model (MCG)
~\cite{bena,deb1} such that $A = 0$ gives generalized model (GCG)
~\cite{bento} and if in addition $\alpha = 1$ one recovers the
original model. Moreover, the first term on the r.h.s. of the
equation (7) gives an ordinary fluid obeying a barotropic equation
of state (EoS) so that we here are essentially dealing with a two
fluid model. Further, the equation (7) points to an EoS that
interpolates between standard fluids at high energy densities and
Chaplygin gas fluids at low energy densities. In the 4D framework
the dynamics of the MCG model has been studied in the reference
~\cite{bena,deb1} and a perturbative study looking for some
generic features is carried out in ~\cite{costa}. On the other
hand Fabris\emph{ et al}~\cite{fab}, in an interesting work, have
used a perturbative analysis to confront observational data within
this model and taking the particular case of power spectrum
observational data have concluded that the recent data restricts
the value of $A < 10^{-6}$ such that the GCG is recovered and the
MCG is almost ruled out. Moreover, in the case of MCG model recent
supernova data seem to favour negative values of the parameter
$\alpha$ ~\cite{deng}. When one attempts to address issues
concerning structure formation the study of cases with negative
values of $\alpha$ becomes more sensible since this implies
imaginary sound velocities, hence plagued with the possibility of
instabilities~\cite{batista, fabris,col}. On the other hand it has
been argued that $\alpha > 1$ is also plausible ~\cite{lu}. With
the help of equations (6) \& (7) a little mathematics shows that
an expression for density comes out to

\begin{equation}
\rho (a) =
a^{-3(1+A)}\left[3(1+A)(1+\alpha)\int\frac{B}{1+A}a^{3(1+A)(1+\alpha)
- 1}~ da + c \right]^{\frac{1}{1+\alpha}}
\end{equation}

where $c$ is an integration constant. The above equation (8)
yields a first integral as
\begin{equation}
\rho = \left[ \frac{B}{(1+A)} +
\frac{c}{a^{3(1+\alpha)(1+A)}}\right]^{\frac{1}{1+\alpha }}
\end{equation}

Plugging in the expression of $\rho$ from equations (3) and (9) we
finally get

\begin{equation}
3 \frac{\dot{a}^{2}}{a^{2}} = \left[ \frac{B}{(1+A)} +
\frac{c}{a^{3(1+\alpha)(1+A)}}\right]^{\frac{1}{1+\alpha }}
\end{equation}
\begin{figure}[ht]
\centering \subfigure[$A = 0.5$]{
\includegraphics[scale=1]{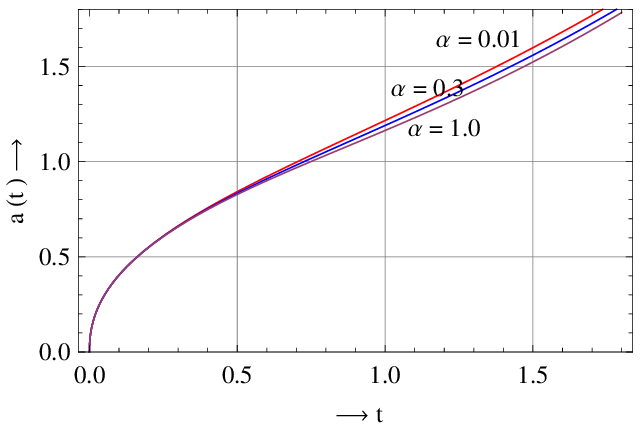}

\label{fig:subfig1} } \subfigure[$A = \frac{1-\alpha}{1+\alpha}
$]{
\includegraphics[scale=1]{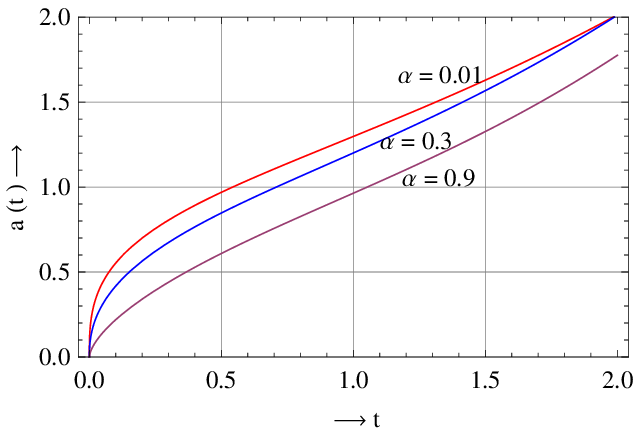}
\label{fig:subfig2} } \subfigure[$A = \frac{1-\alpha}{3(1+\alpha)}
$]{
\includegraphics[scale=1]{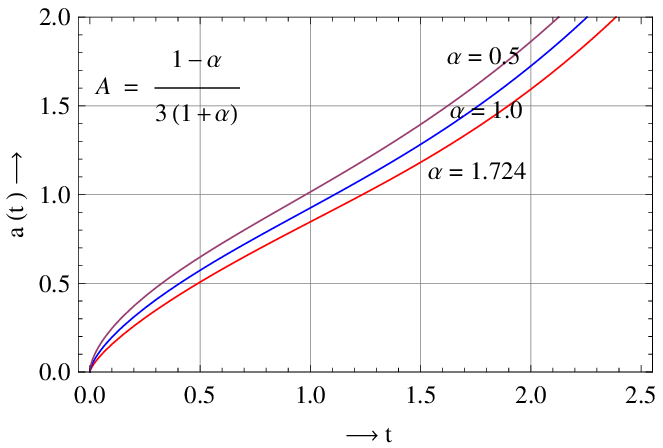}
\label{fig:subfig3} } \label{fig:subfigureExample}
\caption[Optional caption for list of figures]{\emph{The variation of
$a(t)$ and $t$  for different
  values of $\alpha$. The graphs clearly show that
   the scale factor a(t) increases in greater rate for smaller $\alpha$}.}
\end{figure}

 Using equation (10) we have drawn  the figure-1, where  the evolution of $a(t)$ with $t$
 is shown. This figure shows that as $\alpha$ increases the rate of change of scale factor
decreases. A cursory look at the equation also points to this type
of variation of the curves. In fact the equation (10) suggests
that with $\alpha$,
$\frac{\dot{a}}{a}$ becomes flatter.\\

 At this stage if we consider $A = \frac{1-\alpha}{1+\alpha}$ (figure-1b)
we get from the equation (7) that as $\alpha$ increases
($0<\alpha<1$) the first term on the r.h.s. reduces to  zero while
the second term decreases for a particular $\rho$. We get
identical results for $A =
\frac{1-\alpha}{3(1+\alpha)}$(figure-1c). Now, $\rho$ being a
small fraction at the late stage of evolution of the universe any
increase in its value in the exponent finally increases its
magnitude so that the pressure becomes more negative, which, in
turn drives the expansion more vigorously.

Our analysis is based on different sets of observational data. By
using a large sample of milli-arc second radio sources recently
updated and extended by Gurvits et al~\cite{gur} along with the
latest SNeI data as given by Reiss et al~\cite{rei},  Alcaniz and
Lima~\cite{alc} showed that the best fit data for these
observations are $B_{s} = 0.84$ $\&$ $\alpha = 1.0$ (UDME) and
$B_{s} = 0.99$ $\&$ $\alpha = 1.0$ (CGCDM), where
$B_{s}=\frac{B}{\rho_{o}}$, where $\rho_{0}$ is the present
density of the Chaplygin gas. In another work Lima et al
~\cite{lima} showed at $95 \%$ confidence level by the BAO (Baryon
acoustic Oscillation) and Gold sample analysis, the range of
$\alpha$ is $0.9 \leq \alpha \leq 1$ while BAO $\&$ SNLS analysis
provides $0.94 \leq \alpha \leq 1$. Both the results predict
$\alpha$ to be nearly equal to unity. In contrast to this result
Fabris et al ~\cite{fab} as pointed out earlier ruled out the
existence of $A$ for the MCG model in the context of power
spectrum observational data. In this context  the relations $A =
\frac{1-\alpha}{1+\alpha}$ and  $A = \frac{1-\alpha}{3(1+\alpha)}$
seem interesting. When the value of $\alpha$ is nearly equal to
unity there remains a tiny value of $A$, which is not exactly in
line with the work of Fabris et al~\cite{fab}. Lastly Lu et al
~\cite{lu} gives for the MCG best fit data $A = -0.085$ and
$\alpha = 1.724$ in the light of 3 yr WMAP and SDSS data.

 \vspace{0.1 cm}
\section{Cosmological dynamics}

 It is very difficult to get the
exact temporal behaviour of the scale factor, $a (t)$ from the
equation (10) in a closed form because integration yields
elliptical solution only. However, the equation (10) does give
significant information under extremal conditions as briefly
discussed below.

\vspace{0.5 cm}

\textbf{Deceleration Parameter:}\\

 At the early stage of the cosmological evolution
when the scale factor $a(t)$ is relatively small the second term
of the last equation (10) dominates which has been already
discussed in the literature \cite{bena,deb1}. So we will be very
brief on this point. From the expression of the deceleration
parameter, $q$ we get

\begin{equation}
q = -\frac{1}{H^{2}}\frac{\ddot{a}}{a}= \frac{d}{dt} \left(H^{-1}
\right) - 1 = \frac{1}{2} + \frac{3}{2}\frac{p}{\rho}
\end{equation}

where $H$ is the Hubble constant. With the help of the EoS given
by (7) we find

\begin{equation}
q = \frac{1 +3A}{2}- \frac{3B}{2}\frac{1}{\rho^{\alpha +1}}
\end{equation}

which via equation (9) gives

\begin{equation}
q = \frac{1 +3A}{2}- \frac{3B}{2}\left[\frac{B}{1+A}+
\frac{C}{a^{3(1+\alpha)(1+A)}}\right]^{-1}
\end{equation}
\begin{figure}[ht]
\begin{center}
    \includegraphics[width=7.2 cm]{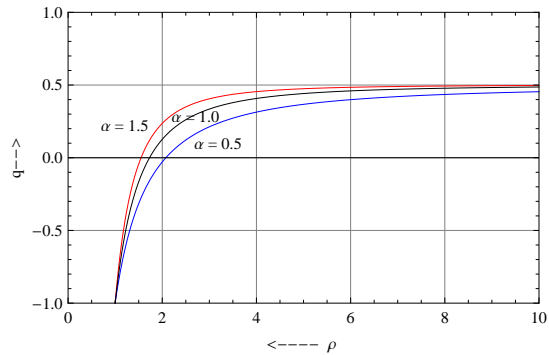}
      \caption{
  \small\emph{The variations of $q$ and $\rho$  for different
  values of $\alpha$. Here $A = 0.001$. }
    }
\end{center}
\end{figure}\\

As the universe expands $\rho$ decreases with time such that the
second term in the equation(12) increases pointing to the
occurrence of a flip when the density attains a critical value
given by $\rho = \rho_{flip} = \left(\frac{3B}{1 +
3A}\right)^{\frac{1}{1 + \alpha}}$. This flip density
$\rho_{flip}$ depends on the exponent $\alpha$ such that at the
larger value of $\alpha$ the flip density decreases, \emph{i.e.},
flip occurs at a lower density, \emph{i.e.}, it occurs at a later
time. We know from Lima's ~\cite{lima} that the value of $\alpha$
is restricted to $0.9 < \alpha < 1$ such that acceleration is a
recent phenomenon. This result is encouraging. As discussed in the
last Section Lu et al ~\cite{lu} argued that $\alpha
> 1$ also conforms to the observational analysis. This finding
is particularly relevant to our case in the sense that higher
values of $\alpha$ signify a lower $\rho_{flip}$, i.e., more
recent accelerating phase. The above analysis in conformity with
the nature of $q \sim \rho$ curve in figure-2.

\vspace{0.5 cm}

\textbf{CASE~A :}~~ At the early stage when the scale factor,
$a(t)$ is very small the  equation (13) reduces to

\begin{equation}
q = \frac{1 +3A}{2}
\end{equation}

Evidently the deceleration parameter has contribution from the
baryonic matter content only such that, $q$  mimcs the ordinary
fluid behaviour with magnitudes $1$ and $\frac{1}{2}$ for
radiation and dust respectively as in a FRW model. When $A=-1$ the
equation (14) gives, $q = -1$  evolving as a $\Lambda CDM$ model.

\vspace{0.5 cm}

 \textbf{CASE~B :} ~~
 In  earlier works  \cite{bena,deb1} authors utilized  the
 above equations to find an equation of state at the late stage of
 evolution as, $ p = \{\alpha +(1 + \alpha)A\}\rho$.

Using the  equations (7) $\&$ (9) straightforward calculations
yield  an \emph{effective} EoS  at the late stage of evolution as

\begin{equation}
p = \rho \left[A - B\rho^{(1+\alpha)}\right]
 = \left[ -1 + \frac{(1+A)^2}{B}
\frac{c}{a^{3(1+A)(1+\alpha)}} \right] \rho = \mathcal{W} (t) \rho
\end{equation}

where

\begin{equation}
\mathcal{W} (t) = -1 + \frac{(1+A)^2}{B}
\frac{c}{a^{3(1+A)(1+\alpha)}}
\end{equation}

which is a function of time only. This is clearly at variance with
the earlier works of ~\cite{ bena,deb1} where the \emph{effective}
EoS shows no time dependence. We also find that at the late stage
of evolution as $a (t) \rightarrow \infty$, $\mathcal{W}
(t)\rightarrow -1 $ so we asymptotically get $ p = - \rho$ from
this Chaplygin type of gas, which corresponds to an empty universe
with cosmological constant such that the equation (11) implies
that the deceleration parameter, $q $ reduces to $-1$.
Interestingly $\mathcal{W}(t)$ always remains greater than $-1$,
thus avoiding the undesirable feature of big rip. In this context
we call attention to a recent work of Z. K. Guo and Y. Z. Zhang
\cite{yz} where a new variant of CG is taken in the form of

\begin{equation}
p = -\frac{B(a)}{\rho}
\end{equation}

where unlike the original CG,  $B$ is taken as a function of the
scale factor $a (t)$. For mathematical simplicity they assumed $B
(a) = B_{0}a^{-n}$ where $B_{0}$ and $n$ are constants and $n <4
$~ and~ $B_{0}>0$. They finally end up with a constant equation of
state parameter

\begin{equation}
w = -1 + n/6
\end{equation}

We find that $n=0$ corresponds to the original Chaplygin gas model
which interpolates between a universe dominated by a dust and
DeSitter era. Moreover $n > 0$ corresponds to a quiessence dominated
 and $n <0$ to a phantom dominated model. In our case we, however, get here a time
 dependent equation of state parameter which always avoids the undesirable
 phantom like behaviour.\\

 To end up a final remark may be in order.
 In an earlier work the present authors ~\cite{dp1,dp11} in the framework of
 $(d + 4)$ homogeneous spacetime studied the scenario with an EoS given by equation(7) but
 generalised to extra dimension. Using an ansatz $b(t) = a(t)^{-m}$
 where $a(t)$ and $b(t)$ are 3D and extra dimensional scale factors and $m$ is
 a constant has led us, at the late stage, to an EoS  $p =
 w \rho$. The expression for the $w$ is found to be
 \begin{equation}
w = - \left[1 + \frac{2dm(m +1)}{k} \right]
\end{equation}

where $k = dm^{2}(d-1) + 6(1-dm)$, is a constant.
 Unlike the usual 4D cases (see for example~\cite{bar}),
 here $w \neq -1 $. Obviously this is due to the presence of extra
 dimensions in the above relation. In 4D case ($d = 0$) $w = -1$ and
 a $\Lambda CDM$ model is the only possibility. In general the magnitude of $w$
is parameter dependent and presents varied possibilities. When $ m
= 0$, \emph{i.e.} $a(t)$ is a constant we again get back the 4D
case. When $m > 0$, $w < -1$; So a phantom like cosmology results
with the occurrence of `big rip' etc. But the cosmology becomes
physically interesting when $-1<m<0$ such that $0> w >-1$ and we
get a quiessence type of model ~\cite{hann,guo2}.

\vspace{0.5 cm}
 The variation of $\mathcal{W} (t)$
with the scale factor  $a(t)$ for different values of $\alpha$ are
shown in the figure-3. We have considered three cases: the
constant value of $A = 0.5$ is chosen for the figure 3a, on the
other hand we have chosen the relation $A =
\frac{1-\alpha}{1+\alpha}$ and $A = \frac{1-\alpha}{3(1+\alpha)}$
for the figure 3b and 3c respectively. All the graphs clearly show
that the scale factor $a(t)$ increases as $\mathcal{W} (t)$
becomes more and more negative.

    Since, $-1\leq \mathcal{W} (t) \leq 0$, the relation shows that $\mathcal{W}
    (t)$ can never be less than $-1$, a good sign. Otherwise there will be
a phantom  stage. In quintessence model $\mathcal{W}
    (t)$ starts from zero and then reduces to  $-1$.
\begin{figure}[ht]
\centering \subfigure[$A = 0.5$]{
\includegraphics[width=6 cm]{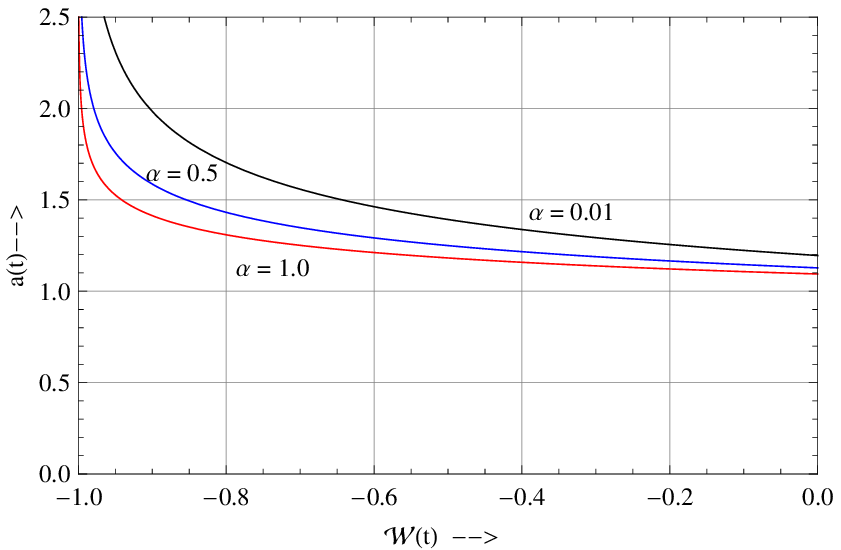}
\label{fig:subfig1} } \subfigure[$A = \frac{1-\alpha}{1+\alpha}$]{
\includegraphics[width=6 cm]{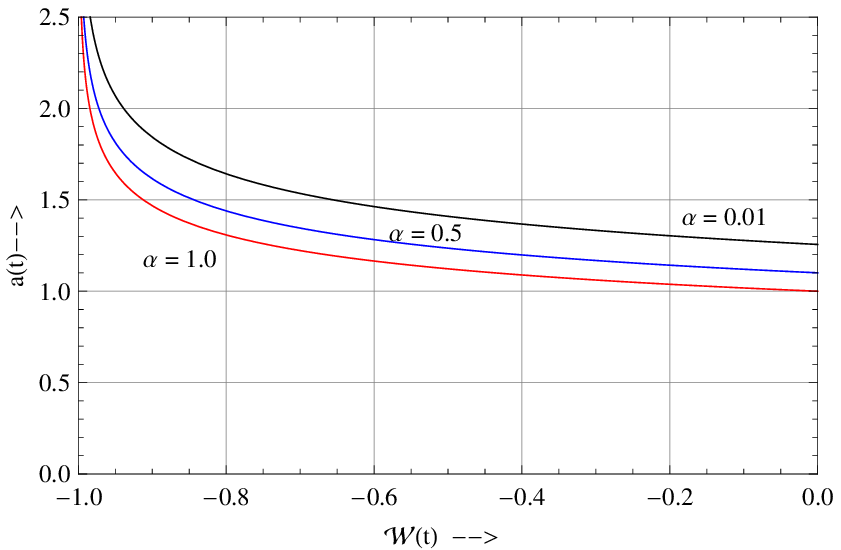}
\label{fig:subfig2} } \subfigure[$A =
\frac{1-\alpha}{3(1+\alpha)}$]{
\includegraphics[width=6 cm]{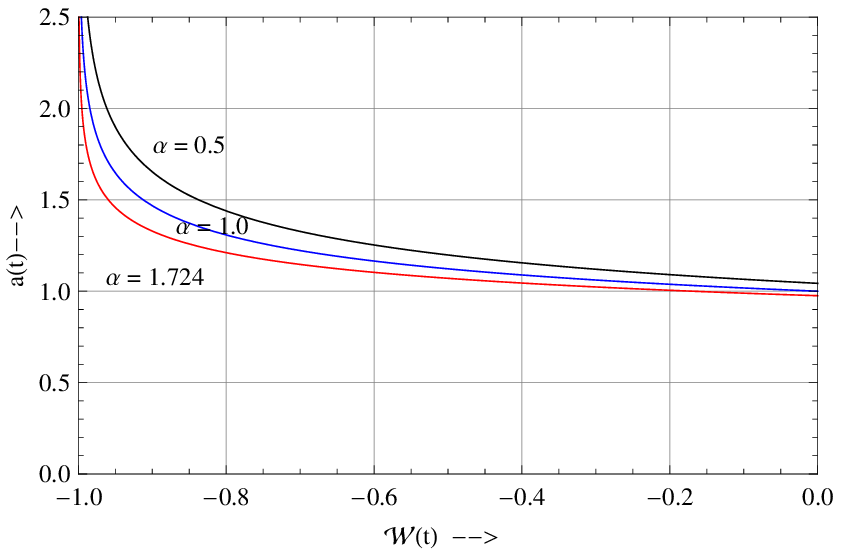}
\label{fig:subfig3} } \label{fig:subfigureExample}
\caption[Optional caption for list of figures]{\emph{The variations of
$\mathcal{W} (t)$ and $a(t)$  for different
  values of $\alpha$. All the  graphs clearly show that the scale factor
    a(t) increases as $\mathcal{W} (t)$ becomes more and more
    negative}.}
\end{figure}
\\

\textbf{Acoustic wave :} \vspace{0.5 cm}

 In this case the
expression of the velocity of sound $v_{s}$  with the help of
equation (15) will be
\begin{equation}\label{c}
v_{s}^{2} = \frac{\partial p}{\partial \rho} = A(1+\alpha) -
\alpha\frac{p}{\rho}= A + \alpha (1+A) \left\{1 -
\frac{c(1+A)}{Ba^{3(1+A)(1+\alpha)}}\right\}
\end{equation}

Using equations (11) and (20) we get the expression of the
deceleration parameters,

\begin{equation}\label{c}
 q = \frac{1}{2} + \frac{3}{2}
\frac{p}{\rho} = \frac{1}{2} + \frac{3A\left(1+\alpha
\right)}{2\alpha} \left\{1 - \frac{v_{s}^{2}}{A\left(1+\alpha
\right)}\right \}
\end{equation}

We have considered three relations for $\alpha$ as $\alpha = 1$,
 $\alpha = \frac{1 - A}{1 + A}$ and $\alpha = \frac{1 - 3A}{1 + 3A}$ to
 study the above equations in a more transparent manner.
 From the observational point
of view it is seen that the value of $\alpha$ is nearly equal to
the unity. As pointed out  earlier  Fabris et al ~\cite{fab}
studied and ruled out the constant $A$. However, our
investigations differ and in a sense more general than Fabris et
al ~\cite{fab} in that we have allowed a small value of $A$ for
$0.9 < \alpha <1$~\cite{lima}. When $\alpha > 1$, we  get the
negative value of $A$ which also is in agreement with some
observational result~\cite{lu}.

\vspace{0.5 cm}

 \textbf{I.}  $\left(\alpha=1\right)$:

From equation (20) we get
\begin{equation}\label{c}
v_{s}^{2} = A +  (1+A) \left\{1 -
\frac{c(1+A)}{Ba^{6(1+A)}}\right\}
\end{equation}
The equation (22) shows that for $A =0 $, $v_{s}$ is always less
than the velocity of light $v_{c}$ and can not be imaginary. For
any other values of A (the limit of $A$ is $0<A<1$) the velocity
of sound $v_{s}$ may not be less than $v_{c}$. Now with the help
of the equation (20) we calculate the condition for $v_{s} \leq
v_{c}$ which is
\begin{equation}\label{c}
a (t) \leq \left[\frac{c \left(1 + A \right)}{B \left \{1-
\left(\frac{1-A}{1+A}\right)\frac{1}{\alpha}\right
\}}\right]^{\frac{1}{3(1+A)(1+\alpha)}}
\end{equation}
 On
the other hand $A \neq 0$ there may be a possibility for $v_{s}
\geq v_{c}$, but our discussion is restricted only to the late
stage of evolution where the scale factor $a(t)$ is large enough.
In this context we get $v_{s}\geq v_{c}$ at the late stage of
evolution. The above phenomenon is shown graphically in the
figure-4a.
\begin{figure}[ht]
\centering \subfigure[$\alpha = 1$]{
\includegraphics[width=6 cm]{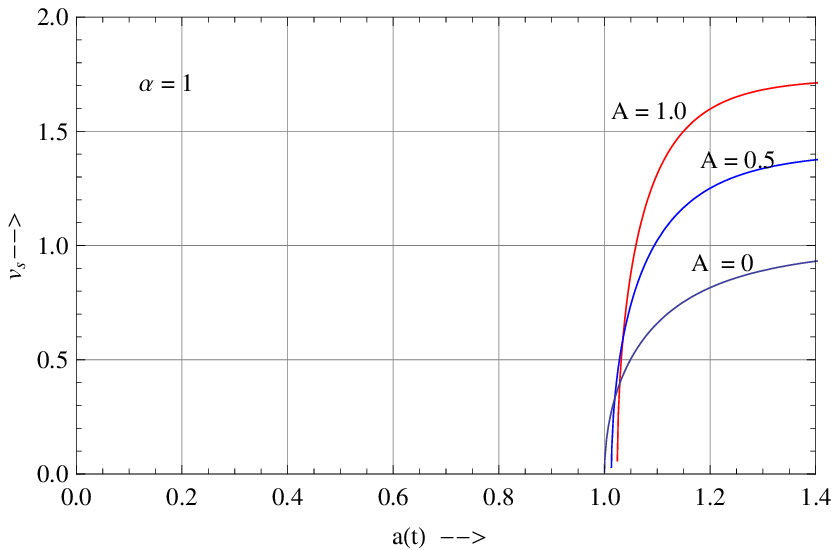}
\label{fig:subfig1} } \subfigure[$\alpha =
  \frac{1-A}{1+A}$ ]{
\includegraphics[width=6 cm]{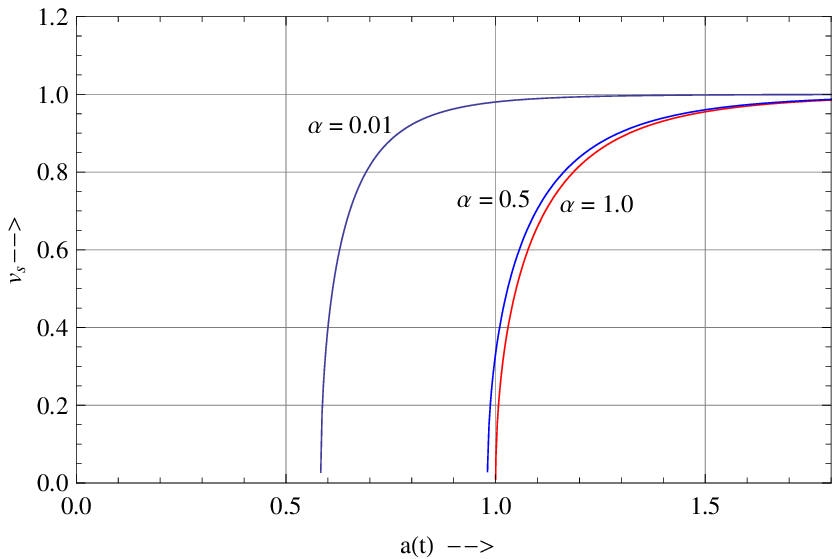}
\label{fig:subfig2} } \subfigure[$\alpha =
  \frac{1-3A}{1+3A}$]{
\includegraphics[width=6 cm]{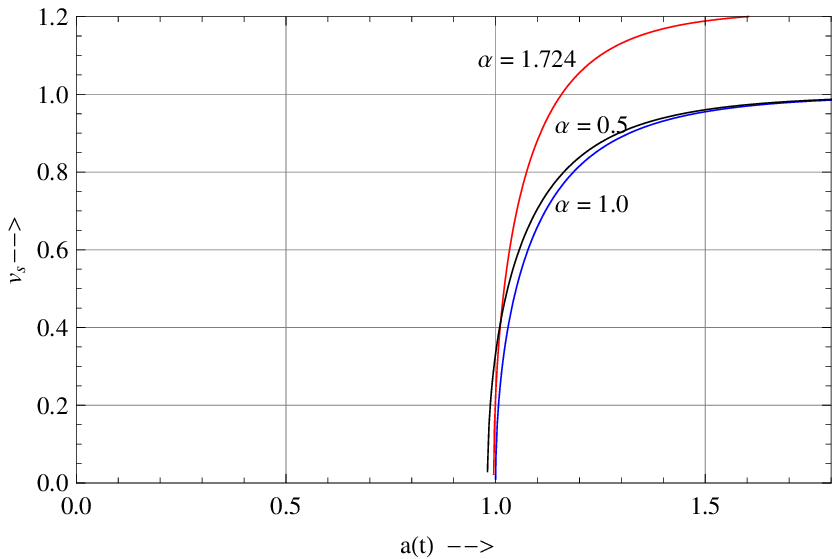}
\label{fig:subfig3} } \label{fig:subfigureExample}
\caption[Optional caption for list of figures]{\emph{The variation of
$v_{s}$ with $a(t)$  for different
  values of $A ~ \& \ ~\alpha$. In the fig.-(a) $v_{s}\leqslant
  v_{c}$  only for A =0, but in the fig.-(b) $v_{s}\leqslant
  v_{c} $  for any value of $\alpha$ (since $0<\alpha<1$) or $A$ (since
$0<A<1$). In the fig-(c), for $  \alpha \leq 1$ gives $v_{s} \leq
v_{c}$ and for $\alpha > 1$ shows $v_{s}>v_{c}$}.}
\end{figure}

\textbf{ II:}  $\left(\alpha=\frac{1-A}{1+A}\right)$: From
equation (20) we get
\begin{equation}\label{c}
v_{s}^{2} = 1 - \frac{4\alpha c}{(1+\alpha)^{2}Ba^{6}}
\end{equation}

 It is evident from the equation (24) as well as from the figure-4b that the velocity of sound $v_{s}$ always less than
the velocity of light $v_{c}$ for any value of $\alpha$ ( since
$0<\alpha<1$). For $\alpha = 1$, $A = 0$ and we get back the
situation depicted in the figure-4a, however, for any other values
of $\alpha$ ($0<\alpha<1$), $A \neq 0$. Some observations predict
that the value of $\alpha$ is nearly equal to $1$ ~ \cite{lima}.
So for the maximum permissible value of $\alpha$, $v_{s}$ should
be always less than the velocity of the light $v_{c}$. But in the
previous case ( for equation (22)), there may be a possibility
that the $v_{s}$ is greater than $v_{c}$~ \cite{gor} for the high
value of $a(t)$ i.e., at the very late stage of evolution. We also
get the same conclusion from the equation (23).

\textbf{ III:}  $\left(\alpha=\frac{1-3A}{1+3A}\right)$:

From equation (20) we get
\begin{equation}\label{c}
v_{s}^{2} = \frac{1-\alpha}{3(1+\alpha)}+\frac{2 \alpha
(2+\alpha)}{3(1+\alpha)}\left\{1 - \frac{2c
(2+\alpha)}{3B(1+\alpha)}\frac{1}{a^{6(2+\alpha)}} \right\}
\end{equation}

For late universe when $a(t)$ is very large, the above equation
reduces to $v_{s}^2 \thickapprox \frac{2 \alpha +1}{3}$. When
$\alpha =0$ i.e., we get back to the $\Lambda CDM$ model, in this
case $v_{s}^2 = \frac{1}{3}$ and for $\alpha = 1$ (in this case $A
= 0$), i.e., for pure Chaplygin gas model, $v_{s}^2 = 1$ which
imply that for $0 <\alpha <1$, $v_{s} \leq c_{s}$. Again for
$\alpha > 1$, we have seen that $v_{s} > c_{s}$, however, violates
the causality condition. The figure-4c gives similar conclusion.\\

Now we discuss the whole analysis of $v_{s}$ in the context of
deceleration parameter $q$ using equation (21). To get
accelerating universe $q$ should be negative. So the condition for
accelerating universe involving $v_{s}$ is $v_{s}^{2} >
\frac{\alpha}{3} + A(1+\alpha)$.
\begin{figure}[ht]
\centering \subfigure[$\alpha = 1$]{
\includegraphics[width=6 cm]{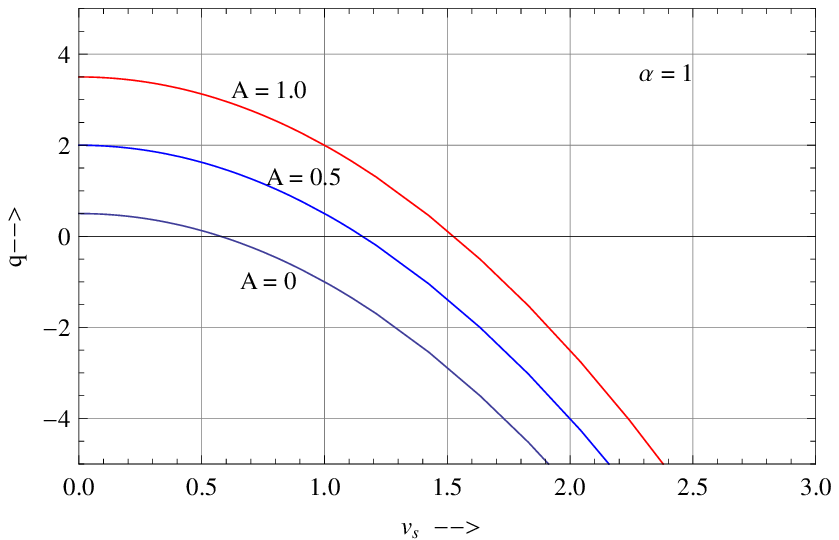}
\label{fig:subfig1} } \subfigure[$\alpha =
  \frac{1-A}{1+A}$ ]{
\includegraphics[width=6 cm]{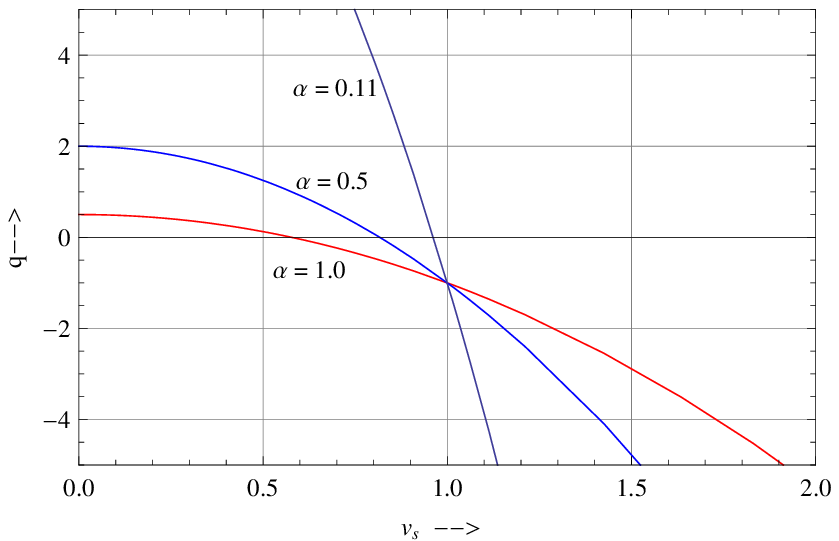}
\label{fig:subfig2} } \subfigure[$\alpha =
  \frac{1-3A}{1+3A}$]{
\includegraphics[width=6 cm]{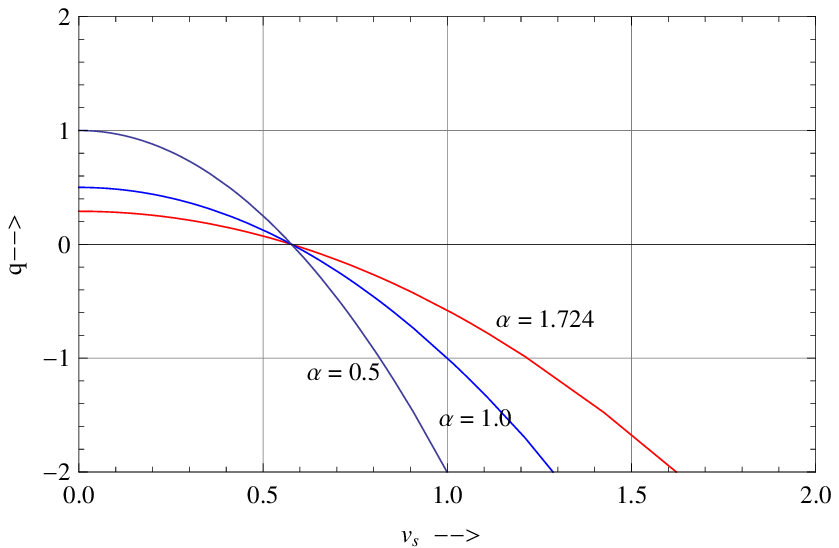}
\label{fig:subfig3} } \label{fig:subfigureExample}
\caption[Optional caption for list of figures]{\emph{The variation
of $q$ with $v_{s}$ for different
  values of $A ~ \& \ ~\alpha$}.}
\end{figure}

From the figure-5a it is seen that flip occurs for $A = 0$ when
$v_{s} < v_{c}$. But for other values of $A$, at the time of flip,
$v_{s} > v_{c}$. In the figure-5b $\&$ 5c at the time of flip,
$v_{s}$ is always less than $v_{c}$ for different values of
$\alpha$ or $A$

Now our analysis will be restricted within the accelerating phase,
\emph{i.e}., after flip. For $\alpha = 1$, to get acceleration
$v_{s}^{2}
> \frac{1}{3} + 2A$. If we consider $A = 0$, \emph{i.e.}, when only
the original  chaplygin type fluid is present, $v_{s}^{2} >
\frac{1}{3}$. So the velocity of sound may or may not be greater
than the velocity of light. For  $A \neq 0$, the above expression
for the velocity of sound further implies that $A \leq
\frac{1}{3}$ in order that $v_{s} < v_{c}$. In this case we
restrict the limit of  $A$ as $0 < A < \frac{1}{3}$. Again when
$\alpha=\frac{1-A}{1+A}$, to get acceleration, $v_{s}^{2} > 1 -
\frac{2 \alpha}{3}$. For $\alpha =1$, $v_{s}^{2} > \frac{1}{3}$
exactly similar to the situation discussed earlier. When $\alpha =
\frac{1 - 3A}{1+3A}$, $v_{s}^2 > \frac{\alpha(1-\alpha)}{9}$. For
$\alpha =0 $ or $1$, $v_{s}^2
> 0$ but for $\alpha >1$, $v_{s}^2 > $ negative value.

 \vspace{0.5 cm}

\textbf{CASE~C :} ~~ Distinctly new models unfold itself when we
take the arbitrary integration constant as $c<0$. Here the energy
density increases with the scale factor mimicing a phantom dark
energy model and finally ending up as a cosmological constant. We
get from (15) that for the matter field to be well behaved the
condition
\begin{equation}
a^{3(1 +A)(1 + \alpha)}> \frac{C(1 + A)}{B(1+\alpha)}
\end{equation}
need to be satisfied. So a minimal value of the scale factor given
by
\begin{equation}
a(t)_{min}= \left[\frac{C(1 +A)}{B(1 +
\alpha)}\right]^{\frac{1}{3(1 +A)(1 + \alpha)}}
\end{equation}

This naturally points to a bouncing cosmology at early times. In
the past Setare ~\cite{set}  analysed these possibilities in a
series of work. To sum up we see that the Chaplygin model
interpolates between a dust at small $a$ and a cosmological
constant at large $a$ but this well formulated quartessence idea
breaks when a negative value of the arbitrary constant is taken.
Following Barrow \cite{bar} if we reformulate the dynamics with a
scalar field $\zeta$ and a potential $V$  to simulate the
Chaplygin cosmology, we find that a negative value of $c$ implies
that we transform $\zeta =i\Psi$. In this case the expressions for
the energy density and pressure corresponding to the scalar field
show
that it represents  a phantom field. \\

 \vspace{0.5 cm}

\textbf{CASE ~D :}

As we are considering a late evolution of our model the last term
in the equation(10) is almost negligible compared to the second
term and so the findings coming from a first order approximation
of the equation(10) may be of relevance.
 Here we find an exact solution of the first order approximation
of the equation(10). Authors of this work are not aware of
attempts of similar kind in any earlier work. So this is clearly a
new result. Now from equation (10) we get, as first order
approximation the equation at the late
 stage of evolution
\begin{equation}
3 \frac{\dot{a}^{2}}{a^{2}}=
\left(\frac{B}{1+A}\right)^{\frac{1}{1+\alpha}} +
\frac{1}{1+\alpha}
\left(\frac{1+A}{B}\right)^{\frac{\alpha}{1+\alpha}}~\frac{c}{a^{3(1+A)(1+\alpha)}}
\end{equation}

For economy of space we skip the intermediate steps and write the
final solution as,

\begin{equation}
\hspace{-0.1 cm} a(t)=\left(\frac{c}{B}\frac{1+A}{1+\alpha}\right
)^{\frac{1}{3(1+A)(1+\alpha)}} \hspace{-0.1 cm}
\sinh^{\frac{2}{3(1+A)(1+\alpha)}}\left
\{\frac{\sqrt{3}}{2}(1+A)^{\frac{1+2\alpha}{2(1+\alpha)}}(1+\alpha)B^{\frac{1}{2(1+\alpha)}}
 \right\}t
\end{equation}

\begin{figure}[ht]
\centering \subfigure[$\alpha = 1$]{
\includegraphics[width=6 cm]{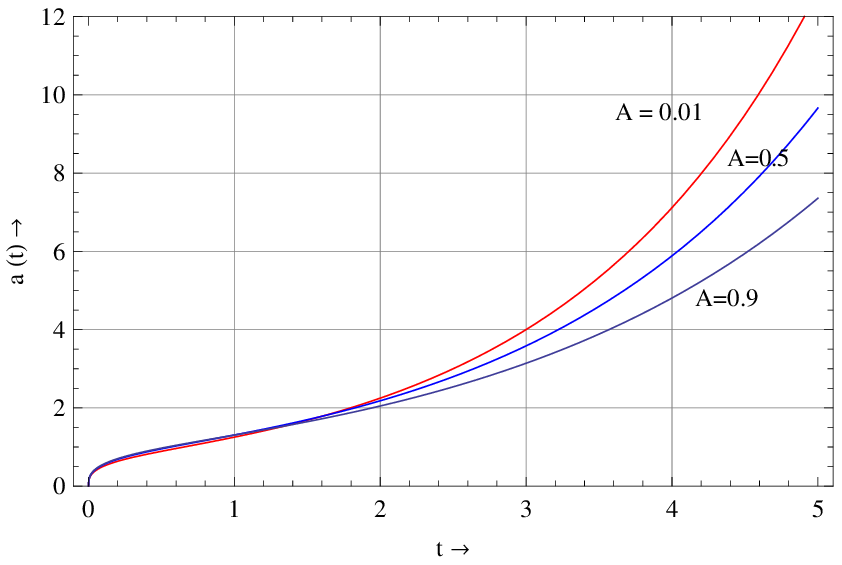}
\label{fig:subfig1} } \subfigure[$\alpha =
  \frac{1-A}{1+A}$ ]{
\includegraphics[width=6 cm]{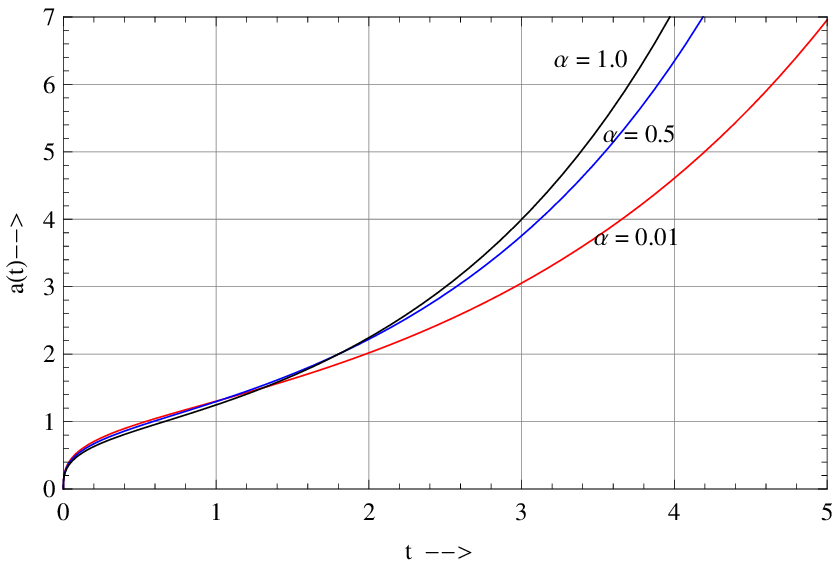}
\label{fig:subfig2} } \subfigure[$\alpha =
  \frac{1-3A}{1+3A}$]{
\includegraphics[width=6 cm]{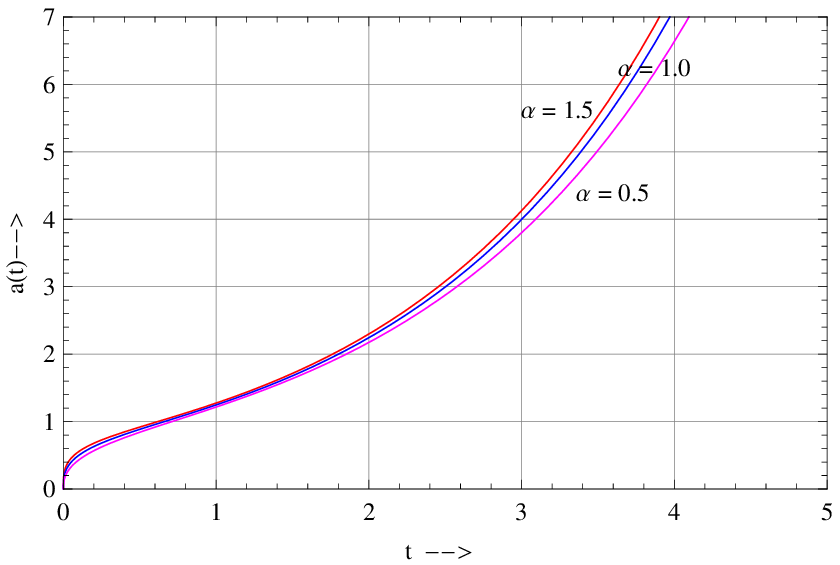}
\label{fig:subfig3} } \label{fig:subfigureExample}
\caption[Optional caption for list of figures]{\emph{The variation
of $a(t)$ with $t$  for different
  values of $A ~ \& \ ~\alpha$  are shown in this figure. The graphs clearly show that the rate of increasing  of the
  scale factor a(t) increases for greater value of
  $\alpha$ or  for smaller value of $A$}.}
\end{figure}

 From figure-6 we have seen that the role of the parameters (A and
 $\alpha$) are just opposite to what we observe in figure-1. A
 plausible explanation may be the fact that unlike the first case
 only first order terms are present here. So the higher order
 terms in the first case drastically change the scenario and makes
 their presence felt in changing the nature of the curves. At this
 stage correspondence to our earlier works \cite{dp1,dp11} may be of
 relevance. We have shown that one may get similar form of
 solution in a higher dimensional spacetime if a particular ansatz
 on the expression of deceleration parameter is taken \emph{apriori
}. But the essential difference between the two lies in the fact
that while in the earlier work the hyperbolic solution results
from a particular form of the deceleration parameter here we have
to invoke a Chaplygin type of gas to get similar comological
evolution.

\vspace{0.5 cm}

 \textbf{  Deceleration Parameter :}

The equation (29) can be reduced in the following form
\begin{eqnarray}
 a(t) = a_{0} \mathrm{sinh}^{n}\omega t
 \end{eqnarray}
 where,
$ a_{0} = \left\{\frac{c}{B (1+\alpha)}\right
\}^{\frac{1}{3(1+A)(1+\alpha)}}(1+A)^{\frac{1+\alpha+\alpha^2}{3(1+A)(1+\alpha)^2}}$,
 $n =  \frac{2}{3(1+A)(1+\alpha)}$ and $\omega =
\frac{\sqrt{3}}{2}(1+A)^{\frac{1+2\alpha}{2(1+\alpha)}}(1+\alpha)B^{\frac{1}{2(1+\alpha)}}$

such that we get from equation (30)
\begin{equation}
q  = - \frac{1}{H^2}\frac{\ddot{a}}{a} =
\frac{1-n~\mathrm{cosh}^{2}\omega t}{n~\mathrm{cosh}^{2}\omega t}
\end{equation}

and
\begin{equation}
t_{c}  = \frac{1}{\omega} \cosh^{-1} \left(\frac{1}{\sqrt{n}}
\right)
\end{equation}
\begin{figure}[ht]

\begin{center}
  \includegraphics[width=7 cm]{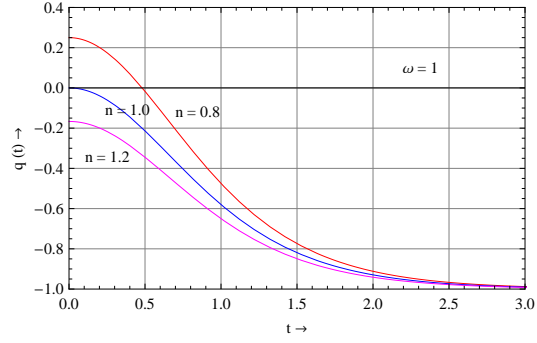}
    \caption{
  \small\emph{The variation of $q$ with $t$  for different
  values of $n$  are shown in this figure. This figures show
  that (i) $n \geq 1 $ gives always acceleration and (ii) $n<1$
  gives flip.
   }\label{1}
    }
\end{center}
\end{figure}

showing that the exponent $n $ critically determines the evolution
of $q $. A little inspection shows that \emph{(i)}  $n \geq 1$
gives always acceleration, \emph{(ii)}   $0< n <1$ gives the
desirable feature of \emph{flip}, although it is not obvious from
our analysis at what value of redshift this \emph{flip} occurs.
Figure-7 gives the similar conclusion that late flip occurs at
lower value of $n$. From equation (31) it further follows that for
physically realistic values of $A$ and $\alpha$ as positive
definite $ 0 < n <1$ and a flip is a distinct possibility, again
it  follows from equation (31) and also from figure-8 that the
early flip occurs at higher values of $n$ as well as $\omega$.
Again $n$ and $\omega$ depend on $A$ and $\alpha$. It can be said
that for constant values of $\omega$ and $A$, late flip occurs at
higher values of $\alpha$, which have some observational
implications that the value of $\alpha$ should nearly equal to
unity $(0.9 < \alpha < 1)$ or greater than unity. If we observe
the expressions of $n$ and $\omega$, we can not say clearly what
values of $\alpha$ and $A$ give the early flip. But we can say
about the time of flip if we consider some special value of
$\alpha$.

\begin{figure}[ht]

\begin{center}
  \includegraphics[width=7 cm]{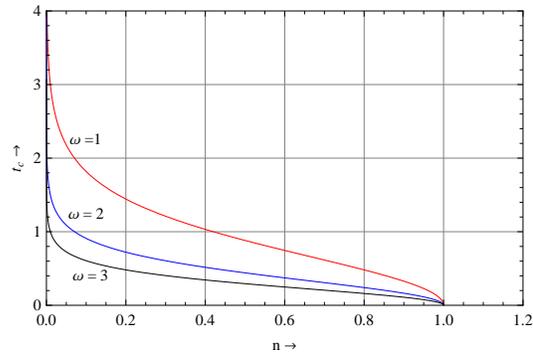}
    \caption{
  \small\emph{The variation of $t_{c}$ with $n$  for different
  values of $\omega$  are shown in this figure. This figure shows
  that the early flip occurs at higher values of $n$ as well as $\omega$.
   }\label{1}
    }
\end{center}
\end{figure}
\vspace{0.5 cm}
Now we consider some special cases.

\textbf{I.}  $\alpha = 1$:

We get from equations (31) $\&$ (32) for $\alpha = 1$ we get the
expression for deceleration parameter

\begin{equation}
q = \frac{3 (1+A) -
cosh^2\left\{\sqrt{3}(1+A)^\frac{3}{4}B^\frac{1}{4} t
\right\}}{cosh^2
\left\{\sqrt{3}(1+A)^\frac{3}{4}B^\frac{1}{4}t\right\}}
\end{equation}

And the flip time $t_{c}$ can be calculated from the above
equation as

\begin{equation}
t_{c} =
\frac{1}{\sqrt{3}(1+A)^\frac{3}{4}B^\frac{1}{4}}~cosh^{-1}\left\{\sqrt{3(1+A)}\right\}
\end{equation}
If $A = -1$, $q = -1$, we get the evolution dominated by $\Lambda$
with no contribution from Chaplygin gas. This also follows from
the equation (34) because here $t_{c} \rightarrow \infty$. Thus
there is no flip as in the de-Sitter model.\\

 \textbf{II.}$\left(\alpha =
\frac{1-A}{1+A} \right)$ :

\vspace{0.2 cm}

Again using equation (31) for $\alpha = \frac{1 - A}{1 + A}$ we
get the expression of the deceleration  parameter
\begin{equation}
q = \frac{3
-\cosh^{2}\left\{\sqrt{3}\left(\frac{B}{1+A}\right)^{\frac{1+A}{4}}t\right\}}
{\cosh^{2}\left\{\sqrt{3}\left(\frac{B}{1+A}\right)^{\frac{1+A}{4}}t
\right\}}=\frac{3
-\cosh^{2}\left[\sqrt{3}\left\{\frac{B(1+\alpha)}{2}\right\}^{\frac{1}{2(1+\alpha)}}t\right]}
{\cosh^{2}\left[\sqrt{3}\left\{\frac{B(1+\alpha)}{2}\right\}^{\frac{1}{2(1+\alpha)}}t
\right]}
\end{equation}

From the equation (32) the flip time $t_{c}$ becomes

\begin{equation}
t_{c} =
\frac{1}{\sqrt{3}}\left(\frac{1+A}{B}\right)^{\frac{1+A}{4}}\mathrm{cosh}^{-1}
\left(\sqrt{3}\right) =
\frac{1}{\sqrt{3}}\left(\frac{2}{B(1+\alpha)}\right)^{\frac{1}{2(1+\alpha)}}\mathrm{cosh}^{-1}
\left(\sqrt{3}\right)
\end{equation}

\textbf{III.}$\left(\alpha = \frac{1-3A}{1+3A} \right)$ :

\vspace{0.2 cm} From equation (31) we get
\begin{equation}
q = \frac{1 - \frac{1+3A}{3(1+A)} \cosh^{2} \left
\{\frac{\sqrt{3}}{2}(1+A)^{\frac{3}{4}(1-A)}\frac{2}{1+3A}B^{\frac{1+3A}{4}}\right
\}}{\cosh^{2} \left
\{\frac{\sqrt{3}}{2}(1+A)^{\frac{3}{4}(1-A)}\frac{2}{1+3A}B^{\frac{1+3A}{4}}\right
\}}
\end{equation}

And from equation (32), the flip time $t_{c}$ becomes

\begin{equation}
t_{c} =
\frac{1+3A}{\sqrt{3}(1+A)^{\frac{3}{4}(1-A)}B^{\frac{1+3A}{4}}}\cosh^{-1}\sqrt{\frac{3(1+A)}{1+3A}}
\end{equation}

\begin{figure}[ht]
\centering \subfigure[$\alpha = 1$]{
\includegraphics[width=6 cm]{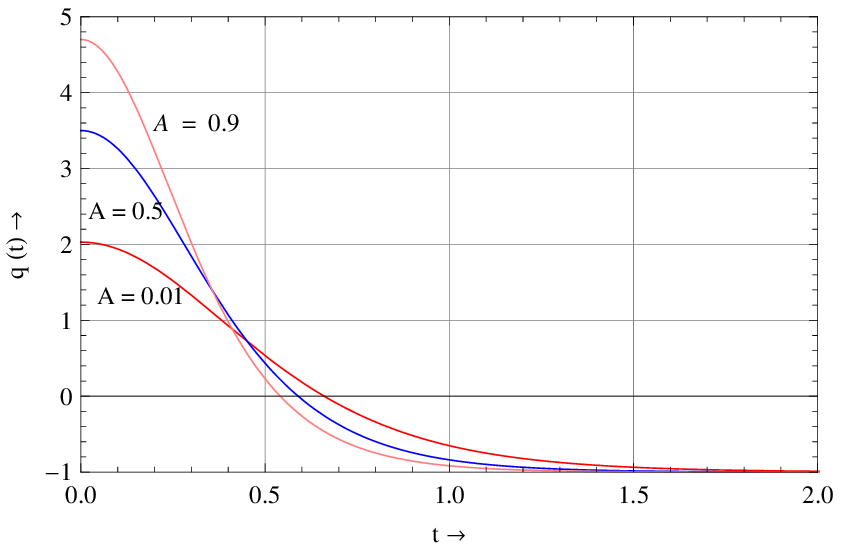}
\label{fig:subfig1} } \subfigure[$\alpha = \frac{1-A}{1+A}$ ]
{\includegraphics[width=6 cm]{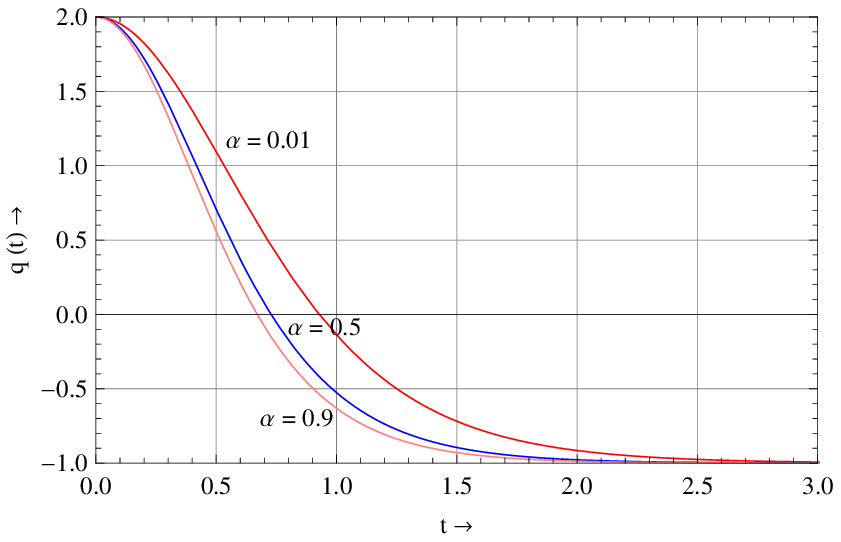} \label{fig:subfig2} }
 \caption[Optional caption for list
of figures]{\emph{The variation of $q$ with $t$  for different
  values of $A ~ \& \ ~\alpha$.
  The graphs clearly show that the flip time  is greater  for
  smaller value of $A$ for the fig.-(a) and increases for greater value of
  $A$ or smaller value of $\alpha$ for the fig.-(b)}.}
\end{figure}

In figure-9, we see that \emph{flip} depends on $\alpha$. From the
graph it is seen that the flip time increase with lower value of
$\alpha$.
 $ a(t)
\rightarrow \infty $, $ t \rightarrow \infty $ in agreement with
our graph.

\section{Raychaudhuri Equation}

It may not be out of place to address and compare the situation
discussed in the last section with the help of the well known Ray
Chaudhuri equation~\cite{ray},  which in general holds for any
cosmological solution based on Einstein's gravitational field
equations. With matter field expressed in terms of mass density
and pressure Ray Chaudhury equation  reduces to a compact form as
\begin{equation}\label{a}
  \dot{\theta}=-2(\sigma^{2}-\omega^{2})-\frac{1}{3}\theta^{2}-\frac{8\pi G}
  {2}\left(\rho+3p \right)
\end{equation}
in a  co moving reference frame. Here $p$ is the
 isotropic pressure and $\rho$ is the energy density from varied sources.
\vspace{0.1 cm} Moreover other quantities are defined with the
help of a unit vector $v^{\mu}$ as under

\begin{subequations}\label{grp}
\begin{align}
\mathrm{the ~expansion ~scalar}~~\theta & =  v^{i}; _{i}  \label{second}\\
 \sigma^{2} & =  \sigma_{ij}\sigma^{ij}  \label{third}\\
\mathrm{the~ shear ~tensor} ~~~~~~\sigma_{ij} &=
\frac{1}{2}(v_{i;j}+ v_{j;i})-
\frac{1}{2}(\dot{v}_{i}v_{j}+\dot{v}_{j}v_{i})-
\frac{1}{3}v^{\alpha}_{;\alpha}(g_{ij} - v_{i}v_{j})  \label{fourth} \\
\mathrm{the~ vorticity~ tensor}~  \omega_{ij} & =
\frac{1}{2}(v_{i;j}- v_{j;i})
-\frac{1}{2}(\dot{v}_{i}v_{j}-\dot{v}_{j}v_{i}) \label{fifth}
\end{align}
\end{subequations}
  We
can calculate  an expression for effective
 deceleration parameter as
\begin{equation}\label{b}
 q= -\frac{\dot{H} + H^{2}}{H^{2}} = -1-3~\frac{\dot{\theta}}{\theta^{2}}
\end{equation}
 which allows us to write,
\begin{equation}\label{c}
  \theta^{2}q = 6 \sigma^{2}+ 12 \pi G \left( \rho
  + 3p \right)
\end{equation}
In our case as we are dealing with an isotropic rotation free
spacetime both  the shear  and vorticity scalars vanish.

With the help of the equations (7), (9) \& (42) we finally get,

\begin{equation}\label{eq:mdiv}
  \theta^{2}q =  12 \pi G
  \left(\frac{B}{1+A}\right)^{-\frac{\alpha}{1+\alpha}}
  \left[-\frac{2B}{1+A}+\frac{c(1+3A+3\alpha+3A
  \alpha)}{1+\alpha}~\frac{1}{a^{3(1+\alpha)(1+A)}}\right]
\end{equation}
In original Chaplygin gas where $\alpha = 1, A = 0$ we get from
the equation (44),  $\theta^{2} q = \frac{2\pi G}{\sqrt{B}}
\left(-B + \frac{c}{a^{6}}\right)$. This is exactly similar to
what we have found in our earlier work \cite{dp2} (vide equation
3.11), when dealing with an inhomogeneous LTB model. Now we
consider some special cases. \vspace{0.5 cm}

\textbf{I.} $\left(\alpha=1\right)$:

 From Equation (44) we get

\begin{equation}\label{c}
  \theta^{2}q =  12 \pi G
  \left(\frac{B}{1+A}\right)^{-\frac{1}{2}}\left[-\frac{2B}{1+A}
  +\frac{c(2+3A)}{a^{6(1+A)}}\right]
\end{equation}

In this case flip occurs when $q = 0$, at that time the scale
factor $a(t)$ will be
\begin{equation}\label{c}
  a(t_{flip}) = \left\{\frac{c}{2B} (1+A)(2+3A)
  \right\}^{\frac{1}{6(1+A)}}
\end{equation}
Now, $q < 0$ at $a(t) > \left\{\frac{c}{2B}
(1+A)(2+3A)\right\}^{\frac{1}{6(1+A)}}$ such that acceleration
takes place in this case.

\textbf{II.} $\left(\alpha=\frac{1-A}{1+A}\right)$:

Again using the equation (44) we get
\begin{equation}\label{c}
  \theta^{2}q =  12 \pi G
  \left\{\frac{B(1+\alpha)}{2}\right\}^{-\frac{\alpha}{1+\alpha}}
  \left[-B(1+\alpha)+\frac{4c}{(1+\alpha)a^{6}}\right]
\end{equation}

 Here, at the flip time $q =0 $ and at that time the
scale factor $a(t)$ will
\begin{equation}\label{c}
 a(t_{flip}) = \left\{\frac{4c}{B}~\frac{1}{(1+\alpha)^2}\right\}
 ^{\frac{1}{6}}= \left[{\frac{c}{B}(A+1)^{2}}\right]^{\frac{1}{6}}
\end{equation}

The acceleration takes place  when $q<0$ \emph{i.e.} $a(t) >
\left[{\frac{c}{B}(A+1)^{2}}\right]^{\frac{1}{6}} $

 \vspace{0.5 cm}

 \textbf{III.} $\left(\alpha=\frac{1-3A}{1+3A}\right)$:\

 From equation (44) we get,
\begin{equation}\label{c}
  \theta^{2}q =  12 \pi G
  \left\{\frac{3B(1+\alpha)}{2(2+ \alpha)}\right\}^{-\frac{\alpha}{1+\alpha}}
  \left[-\frac{3B(1+\alpha)}{(2 + \alpha )}+\frac{2c}{a^{2 (2+\alpha )}}\right]
\end{equation}

Here, at the flip time $q =0 $ and at that time the scale factor
$a(t)$ will
\begin{equation}
a(t_{flip}) =
\left\{\frac{2c}{3B}~\frac{2+\alpha}{1+\alpha}\right\}
 ^{\frac{1}{2(2+\alpha)}}= \left[{\frac{c}{B}(A+1)^{2}}\right]^{\frac{1+3A}{6(1+A)}}
\end{equation}

The acceleration takes place  when $q<0$ \emph{i.e.} $a(t) >
\left[{\frac{c}{B}(A+1)^{2}}\right]^{\frac{1+3A}{6(1+A)}} $

 \vspace{0.5 cm}

In all the  cases discussed above (i.e. for different expressions
of $\alpha$ ), we find out the conditions such that $q < 0$. The
equations (46), (48) and (49) are consistent in the sense that
when $A$ tends to zero
 both the expressions for $a(t_{flip})$ become identical. As discussed in
 the end of the last section the observational constraints point to a tiny value of the constant
 $A$. At this small value of $A$ the expression
  $a(t_{flip})$ in equation (46) is greater than that in
  equation (48). Since $a(t)$ is a monotonically increasing function of time  we get similar
   results from the
  Ray Chaudhury equation also in respect of the flip time which is
  discussed in the previous section
  for small values of $A$. If we consider the equation (49)
    for $\alpha = 1.724$ as Lu's~\cite{lu} choice, in this
    case $A= -0.0886$ (which is close to Lu's data) such
    that $a_{flip} = \left(\frac{0.67c}{B}\right)^{6.8 \times 10^-4}$. \\

\section{Concluding Remarks}

 Here we have considered the homogeneous FRW model with
Modified Chaplygin type gas. Our analysis is based on the results
of different sets of observational data. There is a continued
debate on the exact range of the values of the exponent,$\alpha$
which generalizes the original chaplygin gas. While most
observations point to the value of $\alpha$ as nearly equal to
unity but existing literature abounds with examples of, $\alpha
>1$, which incidentally may give  $v_{s}^{2}
> v_{c}^{2}$. This results in a perturbation of the spacetime and a perturbative
analysis of the whole system shows that it favours structure
formation. While no basic agreement is reached most workers narrow
down the range as  $\alpha$ is $0.9 < \alpha < 1$. Lu et al~
\cite{lu} gives for the MCG best fit data $A = 0.085$ and $\alpha
= 1.724$. In these context we have considered $\alpha = 1$,
$\alpha = \frac{1-A}{1+A}$ and $\alpha = \frac{1 - 3A}{1 + 3A}$
which are in basic agreement with the observational analysis. Our
findings are summarised as follows:

 ~~~~1. As is well known it is very difficult to get exact form of
 solution of the field equations
 so we have studied graphically the variation of scale factor
$a(t)$ with $t$ for different values of $\alpha$. The figure shows
that as $\alpha$ increases the rate of change of scale factor
decreases.

2. We have studied the key equation (10) with the help of
deceleration parameter. From the definition of the deceleration
parameter $q$ we have calculated the flip density $\rho_{flip}$.
At the larger values of $\alpha$ the $\rho_{flip}$ decreases,
i.e., flip occurs at lower density or at a later time. Since the
acceleration is a recent phenomena, this result is in agreement
with the observational analysis that the value of $\alpha$ is
nearly equal to the unity ($\alpha$ is $0.9 < \alpha <1$).  From
the figure-2 it is seen that $\rho_{flip}$ is lower for the higher
values of $\alpha$.

3. Since our universe is accelerating our discussion emphasizes
only  the late stage of evolution. In case B  we get a time
dependent effective equation of state $\mathcal{W}(t)$. It gives
at the late stage of evolution as $a (t) \rightarrow \infty$,
$\mathcal{W} (t)\rightarrow -1 $. So we asymptotically get $ p = -
\rho$ from this Chaplygin type of gas, which corresponds to an
empty universe with cosmological constant such that the equation
(11) implies that the deceleration parameter, $q $ tries to attain
to $-1$. Interestingly $\mathcal{W}(t)$ always remains greater
than $-1$, thus avoiding the undesirable feature of big rip.  Z.
K. Guo and Y. Z. Zhang \cite{yz} considered the new variant of CG
as $B (a) = B_{0}a^{-n}$ where $B_{0}$ and $n$ are constants and
$n <4 $~ and~ $B_{0}>0$. They finally end up with a constant
equation of state parameter. In this case they got the EoS
parameter $w = -1 + n/6$, which is time independent. However, in
our case we can avoid big rip without introducing any extra
parameter.

4. We have studied the velocity of sound in the Modified Chaplygin
Gas model. Here we have discussed the possibility of the speed of
$v_{s}$ is greater than the speed of light. For $\alpha = 1$ and
$A =0$, $v_{s}$ is always less than $v_{c}$, but for $A \neq 0$,
$v_{s}$ exceeds $v_{c}$ at the late stage of evolution. For
$\alpha = \frac{1 - A}{1 + A}$ and $\alpha = \frac{1 - 3A}{1 +
3A}$, we get $v_{s}$ is always less than $v_{c}$.

5. Taking first approximation of the r.h.s. of equation (10) we
get the equation (28). For $\alpha = 1$,  $\alpha =
\frac{1-A}{1+A}$ and $\alpha = \frac{1 - A}{1 + A}$ we get the
solution of equation (28) in the exact form of $a(t) = a_{0}
sinh^{n}\omega t$. We have seen that flip depends upon $\alpha$.
From the figure-7 it is seen that the flip time increases with
lower value of $n$. Moreover the flip time characterized by
equation (36) is found greater than that in equation (34) and
similarly flip time for the equation (38) is greater than the
equation (36). This finding may have some observational
implications. So as $\alpha$ goes to unity, the higher value, $A$
should  vanish. This, however, is in agreement with the Fabris
contention that recent observations point to a vanishing $A$.
Another explanation is that if the value of $\alpha$ is greater
than unity we get the negative value of $A$ as suggested by
Lu~\cite{lu}.

6. The whole exercise is discussed in the context of Raychaudhuri
equation. As expected the results are in broad agreement with the
previous findings.\\

The main drawback of the present analysis is that we have not been
able so far to constrain the model parameters with the help of
observational data as is customary in relevant works in this
field. It would also be a nice idea to use redshifts in place of
cosmic time in most of the equations particularly in drawing the
graphs. That would have been more consistent with the current
nomenclature. Both the issues will be addressed in our future
work.\\

\textbf{Acknowledgments}

 \vspace{0.1 cm}
 DP acknowledges financial support
of ERO, UGC  for a Minor Research project. The financial support
of UGC, New Delhi in the form of a MRP award as also a Twas
Associateship award, Trieste is  acknowledged by SC.


\vspace{0.2 cm}
\begin{thebibliography}{15}
\bibitem{res} Reiss et al, 1998 Astrophys. J. \textbf{116}, 1009, arXiv: 9805201[astro-ph].

\bibitem{spe} D. N. Spergel et al, 2003 Astrophys. J. Suppl. \textbf{148}, 175.

\bibitem{cop} E. Copeland E, M. Sami and S. Tsujikawa, 2006 Int. J. Mod. Phys. \textbf{D15}, 1753.

\bibitem{sam} M. Sami and T. Padmanabhan, 2003 Phys. Rev. \textbf{D67}, 083509, arXiv: 0212317[hep-th].

\bibitem{sch} R. J. Scherrer, 2004 Phys. Rev. Lett. \textbf{93}, 011301.

\bibitem{gib} G. W. Gibbons, 2002 Phys. Lett. \textbf{B 537}, 1.

\bibitem{eli} E. Elizalde, S. Nojiri and S. Odintsov, 2004 Phy. Rev. D \textbf{70}, 043539.

\bibitem{guo} Z. K. Guo, Y. S.  Piao, X. Zhang and Y. Z. Zhang, 2006 Phy. Rev. D\textbf{74}, 127304, arXiv: 0410654[astro-ph].

\bibitem{wan} M. I. Wanas, `Dark Energy: Is It of Torsion Origin?', 2009 Proceedings of the first MEARIM,
edited by A. A. Hady and M. I. Wannas, P-41, arXiv:1006.2154v1
[gr-qc].


\bibitem{neu1} I. P. Neupane, 2009 Class. Quant. Grav. \textbf{26}, 195008, arXiv: 0905.2774
[hep-th].

\bibitem{neu11}I. P. Neupane, 2010 Int. J. Mod. Physics \textbf{D19}, 2281,
arXiv:1004.0254v2 [gr-qc].

\bibitem{dp1} D. Panigrahi and S. Chatterjee, 2011 Grav. Cosm. \textbf{17}, 18, arXiv: 1006.0476v1
[gr-qc].


\bibitem{dp11} D. Panigrahi, Y. Z. Zhang and  S. Chatterjee, 2006 Int.
J. Mod. Phys. \textbf{A21}, 6491, arXiv: 0604079 [gr-qc].

\bibitem{sah} Varun Sahni amd Yuri Shtanov, 2008 `Cosmic Acceleration and Extra
Dimensions' arXiv:0811.3839v1 [astro-ph].

\bibitem{kac} S. Kachru, R. Kallosh,
A. Linde and S. P. Trivedi, 2003 Phys. Rev. D\textbf{68}, 046005,
arXiv: 0301240 [hep-th].

\bibitem{car} M. S. Carroll and L. Mersini,  2001 Phys.
Rev. D \textbf{64}, 124008.

\bibitem{kra}Andrzej Krasi´nski, Charles Hellaby, Krzysztof Bolejko and
Marie-N\"{o}elle C\'{e}l\'{e}rier, 2010 Gen. Rel. Grav.
\textbf{42}, 2453, arXiv: 0903.4070v2

\bibitem{aln} H. Alnes, M Amarzguioui and \O~ Gr\o n, 2007 JCAP \textbf{01}, 007, arXiv: 0506449 [astro-ph].


\bibitem{sc1}S. Chatterjee, 2011 JCAP \textbf{03}, 014.

\bibitem{hir} C. M. Hirata and U. Seljak, 2005 Phys. Rev. \textbf{D72}, 083501 ; arXiv:
0503582[astro-ph].

\bibitem{bent} M. C. Bento, O. Bertolami and A. A. Sen, 2002 Phys.
Rev. D\textbf{66}, 043507.

\bibitem{gor1} V. Gorini, A. Kamenschik and U.
Moschella, 2003,  Phys. Rev. \textbf{D67}, 063509, arXiv: 0209395
[astro-ph].

\bibitem{fabris} J. C. Fabris, S. V. B. Goncalves and P. E.
de Souza, 2002, arXiv: 0207430 [astro-ph]

\bibitem{col} R. Colistete Jr., J. C.
Fabris, S. V. B. Goncalves and P. E. de Souza, 2004 Int. J. Mod.
Phys. \textbf{D13}, 669, arXiv: 0303338 [astro-ph].

\bibitem{bena} H. B. Benaoum, 2002, arXiv: 0205140 [hep-th]

\bibitem{deb1} U. Debnath, A. Banerjee
and S. Chakraborty, 2004 Class. Quant. Grav.\textbf{21}, 5609,
arXiv: 0411015 [gr-qc].


\bibitem{dev1} A. Dev, J. S. Alcaniz and D. Jain,  2003 Phys. Rev. \textbf{D67},
023515, arXiv: 0209379 [astro-ph].

\bibitem{sil} P. T. Silva and O. Bertolami,
2003 Astrophys. J. \textbf{599}, 829, arXiv: 0303353[astro-ph].

\bibitem{dev2} A. Dev, D. Jain and J. S. Alcaniz, 2004 Astron. Astrophys
\textbf{417}, 847, arXiv: 0311056 [astro-ph].

\bibitem{alcaniz} J. S. Alcaniz, D. Jain and A. Dev, 2003  Phys.
Rev. \textbf{D67}, 043514, arXiv: 0210476 [astro-ph].

\bibitem{bento} M. C. Bento, O. Bertolami and A. A. Sen, 2003 Phys.
Rev.\textbf{D67}, 063003, arXiv: 0210468 [astro-ph].

\bibitem{cunha} J. V. Cunha, J. S. Alcaniz and J. A. S. Lima, 2004
Phy. Rev. \textbf{D69}, 083501, arXiv: 0306319[astro-ph].

\bibitem{sahni} V. Sahni, T. D. Saini and A. A. Starobinsky and U.
Alam, 2003  JETP Lett. \textbf{77}, 201; arXiv: 0210498
[astro-ph].

\bibitem{dp2} D. Panigrahi and S. Chatterjee, 2011 JCAP \textbf{10}, 002, arXiv: 1108.2433 [gr-qc].

\bibitem{fab1} J. C. Fabris and J. Martin, 1997  Phys. Rev.
\textbf{D55}, 5205.

\bibitem{lima} J. A. S. Lima, J. V. Cunha  and J. S. Alcaniz,
2008 Astropart. Phys \textbf{30},196, arXiv: 0608469 v1
[astro-ph].

\bibitem{costa} S. Costa, M. Ujevic and A. F. dos Santos, 2008 Gen.
Rel. Grav. \textbf{40}, 1683.

\bibitem{fab} J. C. Fabris, H. E. S. Velten, C. Ogouyandjou and J.
Tossa, 2011 Phys. Lett \textbf{B694}, 289, arXiv: 1007.1011v1
[astro- ph.CO].

\bibitem{deng} Xue-Mei Deng, 2011 Braz. J. of Physics \textbf{41}, 333, arXiv: 1110.1913v1 [gr-qc].

\bibitem{batista} C. E. M. Batista, J. C. Fabris and M. Morita,
2010 Gen. Rel. and Grav. \textbf{42},839, arXiv: 0904.3948
v1[gr-qc].

\bibitem{lu} Lu et al, 2008 Physics Lett \textbf{B 662}, 87,
  arxiv: 1004.3364 [astro-ph].


\bibitem{gur} L. I. Gurvits, K. I. Kellermann and S. Frey, 1999 Astron. and Astrop. \textbf{342}, 378, arXiv: 9812018 [astro-ph].

\bibitem{rei} A. Riess et al, 2004  Astrophys. J. \textbf{607}, 665, arXiv: 0402512 [astro-ph].

\bibitem{alc} J. S. Alcaniz and J. A. S. Lima, 2005  Astrophys. J. \textbf{618 }
16, arXiv: 0308465v2 [astro-ph].

\bibitem{yz} Z. K. Guo and Y. Z. Zhang, 2007 Phys. Lett. \textbf{B 645}, 326, arXiv: 0506091 [astro-ph].

\bibitem{bar} J. D. Barrow, 1988 Nucl. Phys. \textbf{B 310}, 743.

\bibitem{hann}  S. Hannestad and E. M\"{o}rtsell, 2002 Phys. Rev. \textbf{D66},
063508.

\bibitem{guo2} Z. K. Guo, N. Ohta and Y. Z. Zhang, 2005 Phys. Rev.
\textbf{D72}, 023504, arXiv: 0505253 [astro-ph].


\bibitem{gor} V. Gorini, A. Yu Kamenshchik, U. Moschella, O. F. Piattella
and A. A. Starobinsky, 2009 Phys. Rev. \textbf{D80}, 104038,
arXiv: 0909.0866v3 [gr-qc].


\bibitem{set} M. R. Setare, 2007 Eur. Phys. J.  \textbf{C52}, 689,
arXiv: 0711.0524 [gr-qc].

\bibitem{ray} A. K. Raychaudhuri, 1955 Phys. Rev. \textbf{98},
1123.
\end{thebibliography}
\end{document}